\documentclass[aps,prl,amsmath,twocolumn,amssymb,titlepage,superscriptaddress]{revtex4-1}
\usepackage{graphicx}
\usepackage{nicefrac}
\usepackage{amsfonts}
\usepackage{amssymb}
\usepackage{amsmath} 
\usepackage{subfigure}
\usepackage{color}
\usepackage{multirow} 
\usepackage{tabularx} 
\usepackage{array}
\usepackage{units}
\usepackage{tensor} 
\usepackage{braket}
\usepackage{bm}
\usepackage{hyperref}
\usepackage{soul}
\usepackage[normalem]{ulem}

\emergencystretch=\maxdimen
\begin{document}

\title{Pressure Effects on the 4f Electronic Structure of Light Lanthanides}

\author{W.-T. Chiu}
\affiliation{Physics Department, University of California, Davis,
California 95616, USA}
\author{D. R. Mortensen}
\affiliation{Department of Physics, University of Washington, Seattle,
Washington 98195-1560, USA}
\author{M. J. Lipp}
\affiliation{Physics Division, Lawrence Livermore National Laboratory,
Livermore, California 94550, USA}
\author{G. Resta}
\affiliation{Physics Department, University of California, Davis,
California 95616, USA}
\author{C. J. Jia}
\affiliation{Stanford Institute for Materials and Energy Sciences, SLAC
National Accelerator Laboratory, Menlo Park, California 94025, USA}
\author{B.~Moritz}
\affiliation{Stanford Institute for Materials and Energy Sciences, SLAC
National Accelerator Laboratory, Menlo Park, California 94025, USA}
\author{T. P. Devereaux}
\affiliation{Stanford Institute for Materials and Energy Sciences, SLAC
National Accelerator Laboratory, Menlo Park, California 94025, USA}
\affiliation{Geballe Laboratory for Advanced Materials, Stanford
University, Stanford, California 94305, USA}
\author{S. Y. Savrasov}
\affiliation{Physics Department, University of California, Davis,
California 95616, USA}
\author{G. T. Seidler}
\affiliation{Department of Physics, University of Washington, Seattle,
Washington 98195-1560, USA}
\author{R. T. Scalettar}
\affiliation{Physics Department, University of California, Davis,
California 95616, USA}

\date{\today}

\begin{abstract}
Using the satellite structure of the $L\gamma_1$ line in non-resonant
x-ray emission spectra, we probe the high-pressure evolution of the bare
4f signature
of the early light lanthanides at ambient temperature. For Ce
and Pr the satellite peak experiences a sudden reduction concurrent
with their respective volume collapse (VC) transitions.  
These new experimental results are supported by calculations using
state-of-the-art extended atomic structure codes for Ce and Pr, and
also for Nd, which does not exhibit a VC. 
Our work suggests that changes to the 4f occupation are
more consistently associated with evolution of the satellite than 
is the reduction of the 4f moment.
Indeed, we show that in the case of Ce, mixing of a higher atomic
angular momentum state, driven by the increased hybridization, acts to obscure the
expected satellite reduction.
These measurements
emphasize the importance of a unified study of a full
set of microscopic observables to obtain
the most discerning test of the underlying, fundamental f-electron
phenomena at high pressures.
\end{abstract}
\maketitle

\underbar{Introduction:}
The physics and chemistry of lanthanides is of critical importance to fields from
catalysis,\cite{edelmann12,kilbourn86,shibasa02}, separations chemistry
of nuclear waste,\cite{NucWaste} to cuprate
superconductivity\cite{bednorz86,ivanovskii08} and
bioscience.\cite{chen06,bouzigues11} Despite a rich history, theoretical treatment of these materials remains a
fundamental challenge. The difficulty stems primarily from the
underlying nature of f-electron states. In materials with partially
filled f shells, the electrons occupy narrow strongly
correlated energy bands. The resulting electronic
interactions exist between the well understood atomic and uncorrelated
band limits and are responsible for a veritable zoo of exotic behaviors:
metal-to-insulator transitions,\cite{imada98}
superconductivity,\cite{pfleiderer09} hidden orderings,\cite{chandra02}
etc. 

A primary example of the challenges in modeling emergent f-electron phenomena 
is the volume collapse (VC): at high pressures
several lanthanides undergo a first-order phase transition that results
in large changes to lattice constants and
resistivity.\cite{egorov97,velisavljevic04} 
After the discovery of this phenomenon in Ce,
\cite{bridgman27,bridgman48,lawson49} 
attention focussed on differentiating
the Hubbard-Mott (HM)\cite{johansson74,johansson75} picture,
which considers the increasing interatomic overlap of 4f orbitals with
pressure,
and Kondo volume
collapse (KVC)\cite{allen82,lavagna82,allen92,mcmahan03,held01,haule05} 
scenario,
where the screening of the 4f electrons by broad conduction bands is
paramount.

While both HM and KVC theories have had notable successes for Ce,
predicting macroscopic observables such as the Ce equation of state,
\cite{lipp08,johansson09} 
they are distinguished by contrasting
expectations for the behavior of a foundational microscopic observable:
the presence of a 4f localized electron and related properties such as
the magnetic moment.  Upon crossing the VC transition, the complete 4f
itinerancy predicted by the HM model necessarily extinguishes the 4f
moment whereas the hybridization, leading to increased f-screening, of
the KVC anticipates smaller effects.  
A significant step forward was the recognition of
similarities, for example in the evolution with pressure of the density of states, between the two viewpoints\cite{held01,zolfl01}.

Motivated by these observations, Lipp {\it et al.}~presented a study of
the pressure evolution of the $L\gamma_1 (4d_{3/2} \rightarrow
2p_{1/2}$) non-resonant x-ray emission spectra (NXES) for metallic Ce
across its VC.\cite{lipp12} Analogous to the $K\beta'$ feature in
3d-transition metal NXES, the $L\gamma_1$ line exhibits a lower-energy
satellite feature, $L\gamma_1'$, arising from (4f, 4d) exchange. As a
result, the emission intensity is highly sensitive to
the presence of a 4f electron,  
raising several fundamental unresolved questions concerning
the relative importance of the f occupancy and the moment in the 
evolution of the $L\gamma_1'$ satellite.

Here we report an extension of the Lipp {\it et al.}
work to La, Pr, and Nd trivalent metals which lends further insight
into the simultaneous occupation, moment, satellite evolution with 
hybridization. 
Specifically, we compare the volume-collapsed NXES data to
predictions made by advanced atomic calculations, employing
LDA-determined values for the hybridization.
The chief conclusion of this paper is
that it is the 4f {\it occupation}, rather than the moment,
which more consistently tracks the intensity of
the $L\gamma_1'$ satellite.  
This opens the door to 
interpretation of NXES spectroscopy of these materials in which the 4f
occupation plays a more central role.


\underbar{Experimental Results:}
We present the measured $L\gamma_1$ NXES spectra as a function of
pressure for La, Ce, Pr, and Nd in Fig.~1. For clarity,
the data in Fig.~1(a) has been smoothed using a non-broadening 
third-order Savitzky-Golay filter with a 5 eV window (comparable to the
lifetime-limited resolution of 3.7-4.0 eV) and normalized to peak
intensity. Raw data (without smoothing) is used in Fig.~2 and
analysis below. Also shown is the highest pressure data subtracted from
lowest pressure for each element, to highlight the
influence of pressure on the spectral shape. 
In Fig.~1(b) we present a quantitative extraction of the relative
$L\gamma_1'$
intensities. While comparison to La is a useful diagnostic for Ce as
shown above, it does not work well for Pr and Nd where the differences
in the multiplet structure of the $L\gamma_1$ peak prevent
a direct cross evaluation. Instead we fit the spectrum to a
sum of Lorentzians:  one to model the contribution from the main peak
in the satellite region and the others (two for Pr and three for Nd) to fit the $L\gamma_1'$ peak
proper. This procedure has previously been used for Ce
NXES.\cite{lipp12} The
results shown in Fig.~1(b) reflect the spectral weight of the shoulder
relative to the main peak. For Ce, Pr, and Nd these results have been
normalized to the largest intensity below the transition for each
sample to eliminate small run-to-run
variations due to differences in spectrometer tune up that would
otherwise prevent direct comparison.

La, which was found to have constant, negligible satellite intensity at all
pressures, is left unnormalized. Nd is similarly found to have
negligible variation even during the minor change in spectral shape upon
the fcc to distorted-fcc transition at 18.0 GPa discussed below.
Further measurements are required to determine if Nd eventually
undergoes $L\gamma_1'$ reduction, similar to Ce and Pr, upon reaching
its delocalized $\alpha-U$ phase at $\sim 100$ GPa.\cite{chesnut00} 
Before interpreting these results,
however, some brief context is needed on known behavior of the light
lanthanides under pressure and on the underlying physics of the
$L\gamma_1$ NXES spectra.

\begin{figure}
\includegraphics[width=1.77in]{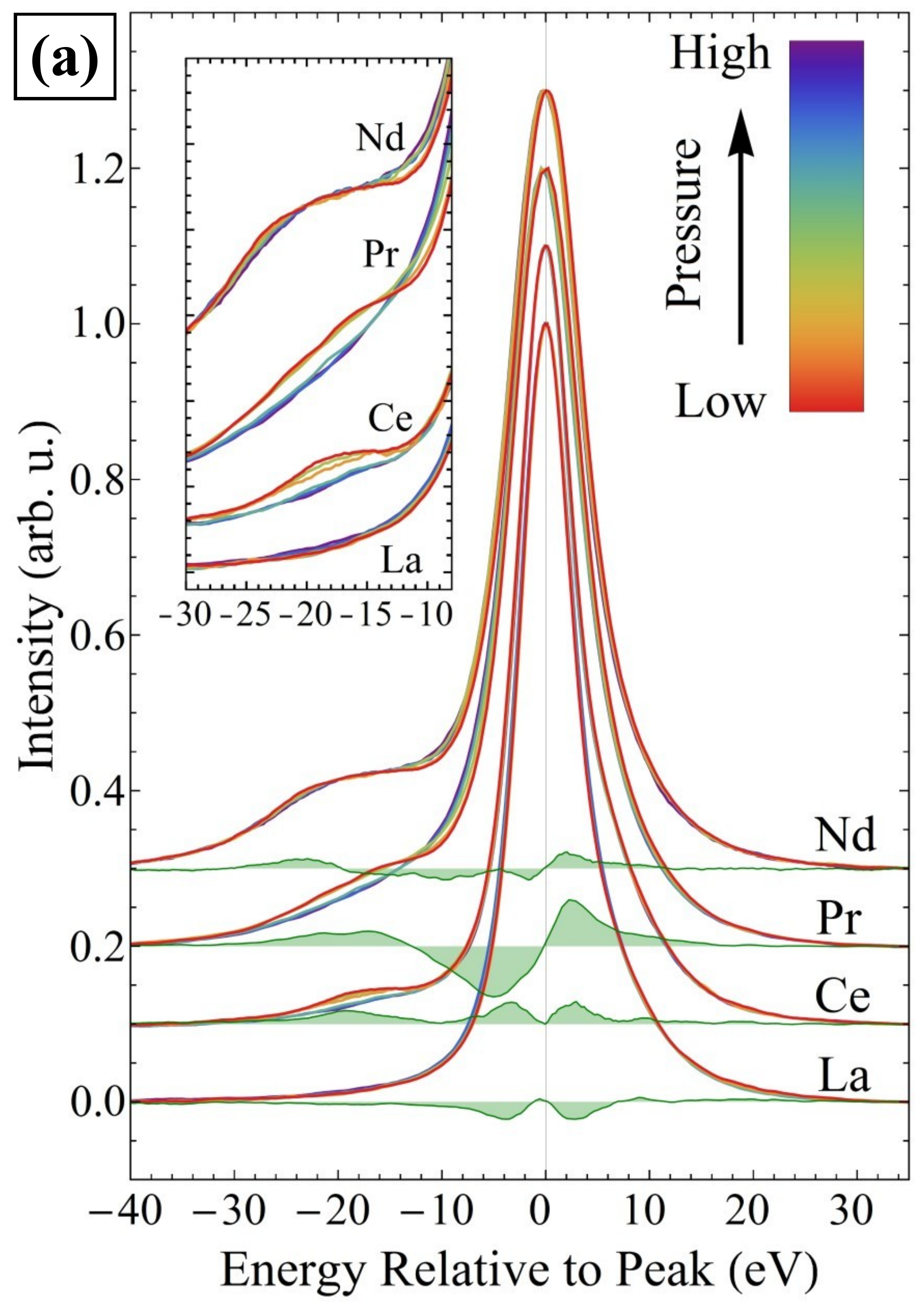}
\includegraphics[width=1.58in]{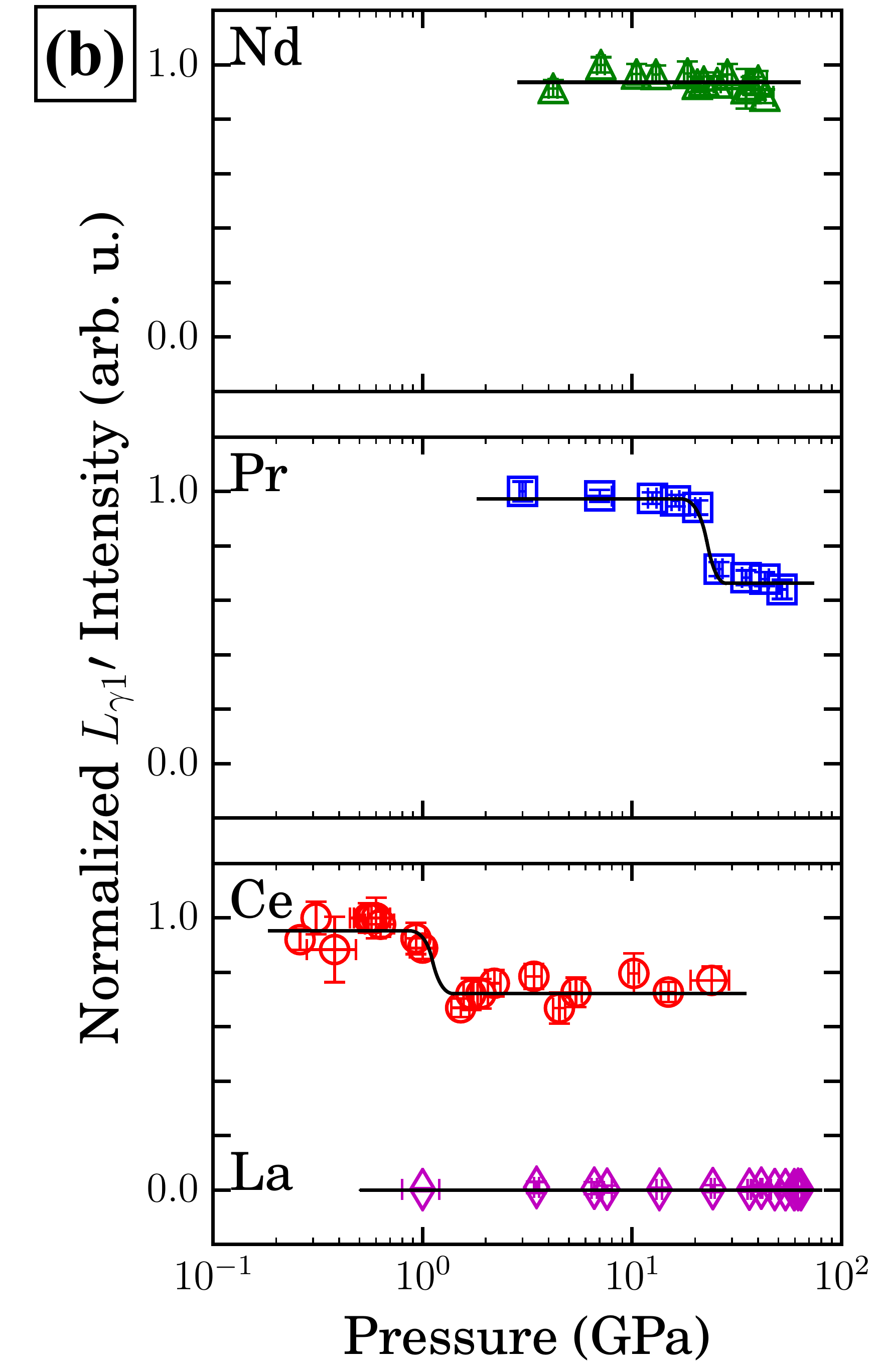}

\caption{(Color online.) 
(a) Experimental $L\gamma_1$ spectra normalized to peak
intensity. A non-broadening, third-order Savitzky-Golay smoothing filter
of 5-eV window is applied to each spectrum.  For clarity, each
element has been offset by 0.1. Also shown is the highest pressure data
subtracted from the lowest pressure data for each element (shaded
green).
(b) Experimental intensity of the $L\gamma_1'$ satellite 
calculated relative to the main $L\gamma_1$ peak. For Ce, Pr, and Nd
the intensities have been normalized to the largest value below the
transition for each sample in order to reduce the effects of run-to-run
variations. La, which has negligible $L\gamma_1'$ intensity, is left
unnormalized. The solid black lines are guides for the eye. 
}
\end{figure}

First, under pressure the lanthanide crystal structures initially pass
through several high-symmetry transformations of
different stacking sequences of close-packed layers, later transitioning
to low-symmetry early-actinide-like phases indicative of
f-electron bonding.\cite{holzapfel05,mcmahan98,lindbaum03} 
While in Pr this transition is accompanied
with a large VC ($\sim$10\%),
\cite{olsen85,vohra99,mao81,smith82,baer03}
Nd reaches its low-symmetry
$\alpha-U$ structure entirely through smooth 
transformations.\cite{chesnut00} Hence,
Nd does not exhibit a VC transition. Ce is a unique case as it
experiences an iso-structural (fcc) VC (15\%),
unassociated with the high-to-low symmetry transition.  For this reason
its VC is thought to be primarily electronically-driven.

\begin{figure*}
\centering
\includegraphics[width=6.6in]{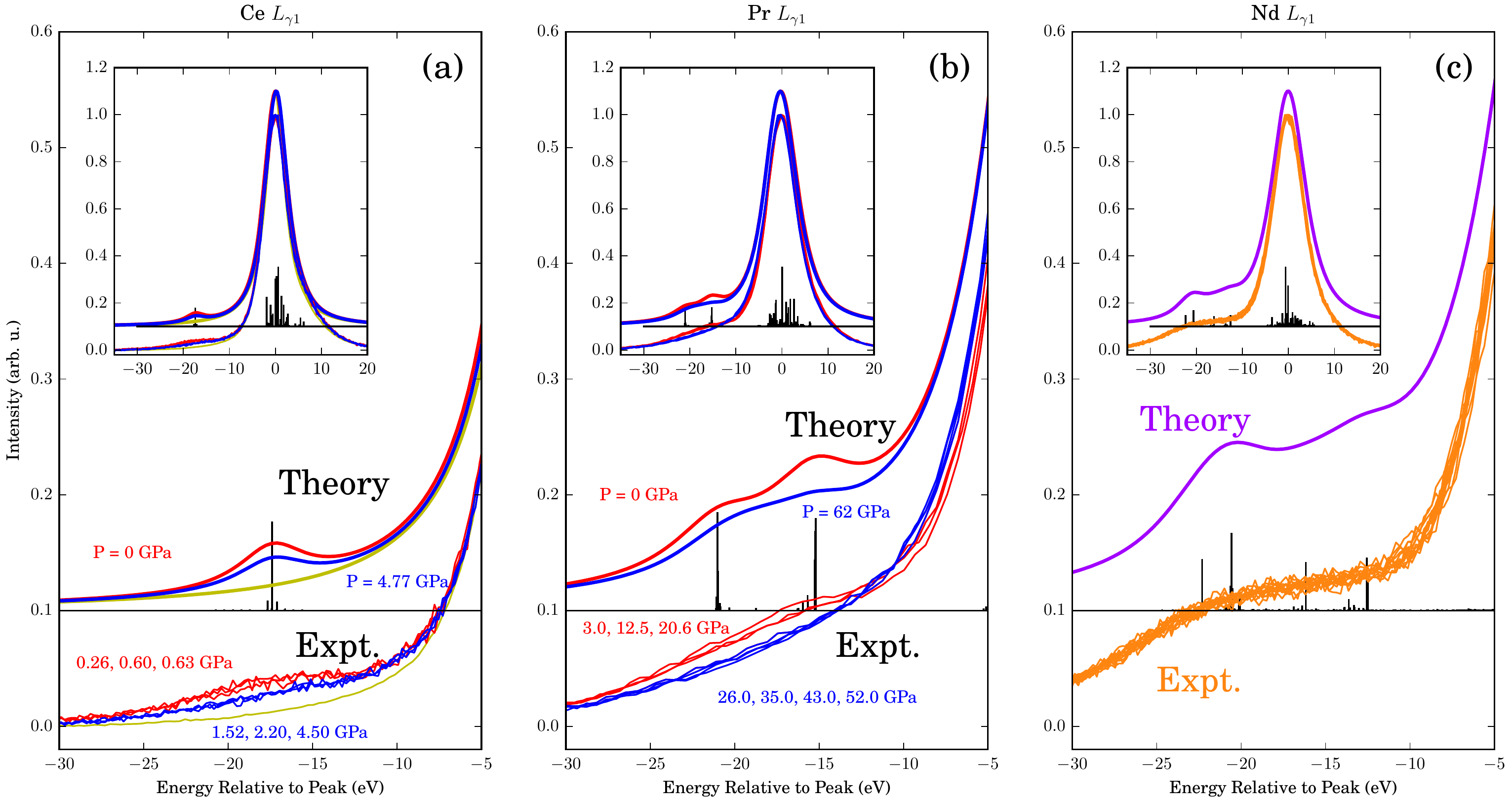}
\caption{(Color online.) The shoulder peak (satellite) 
region of the experimental
and calculated X-ray emission spectra for light lanthanide metals.
The red(blue) curves are
at low(high) pressure.  In the theory 
spectra, low pressure is modeled using parameters at P=0 GPa, and the blue 
curves are obtained using high pressure parameters. (4.77 GPa for Ce and 62 GPa for Pr.) The
yellow curves in (a) are results of lanthanum for a zero shoulder peak
reference. (c) The Nd $L\gamma_1$ spectra do not change with
pressure up to 43 GPa.  Here experimental
and theoretical results are shown as orange and purple respectively.
The calculated spectra are shifted by +0.1. 
Insets show the full $L\gamma_1$ spectra. The vertical black lines
are the calculated transition probabilities before broadening at zero pressure. }
\end{figure*}

Second, as demonstrated by Lipp {\it et al.},
\cite{lipp12} NXES provides a
sensitive measure of the evolution
of intrinsic 4f signatures in Ce. In
$L\gamma_1$ NXES a high-energy photon, tuned well above the $L_2
(2p_{1/2})$ edge, promotes a $2p$ electron into the continuum. The
resulting core-level vacancy is unstable and may be filled by a
$4d_{3/2}$ electron accompanied by either photon emission 
($L\gamma_1$ x-ray fluorescence) or Auger electron ejection.
When the lanthanide species has a nonzero 4f-occupancy, a low energy
satellite ($L\gamma_1'$) appears below the main
$L\gamma_1$ fluorescence peak due to intra-atomic exchange between
4f and 4d orbitals.\cite{glatzel05} The
relative intensity and position of the $L\gamma_1'$ shoulder reflects
the strength of the coupling and is directly sensitive to 
the 4f properties.\cite{jouda97, Tanaka95}

These observations give rise to a fundamental question:
To what extent is the evolution of the satellite a probe 
of the bare moment or of the occupation?
This question complements the well-considered debate concerning whether the VC
itself is associated with destruction of the bare moment or its
screening.
Of course, 
the moment is linked to the occupation, so the answer is not expected to
be completely crisp.  In the  theoretical work to follow we will compute
the satellite peak, occupation, and bare moment as functions of
hybridization (pressure) to lend insight into these issues.

With these details in mind, we return to Fig.~1. First, note that
the La $L\gamma_1$ NXES spectra show no $L\gamma_1'$ exchange peak up to
$64.0 \pm 3.0$ GPa, beyond the reentrant fcc phase starting at 60 GPa,
\cite{porsch93}
indicating no change from its nominally $4f^0$ configuration. This null result indicates that the changes observed in Ce, Pr and Nd $L\gamma_1'$ peaks are physically meaningful. That being said, La does display an apparent
broadening of the main $L\gamma_1$ peak (on the order of $\sim 0.5$ eV)
with increased pressure. We propose that this effect
is due to increased
splitting in the multiplet structure underlying to the $L\gamma_1$ peak.
The changing spectral shape therefore likely contains valuable
information on the evolving 4d-electron interactions
\cite{degroot05}  and thus merits
future theoretical consideration.

Second, in contrast to La ($4f^0$), Ce and Pr, which are nominally $4f^1$
and $4f^2$ at ambient conditions, exhibit large and sudden decreases in
$L\gamma_1'$ intensity concurrent with the VC transitions (0.9 GPa and
20.0 GPa respectively). Taken naively this result could be used as
evidence in support of the HM model as described above. However, it must be
noted that although the increased 4f-5d hybridization predicted
by KVC invariably mixes the 4f electrons out of their native orbitals,
leading to deviations from the ground state electronic configuration;
the hybridization causes a rise in the $f^{n\pm 1}$ configuration
weights at the expense of the sharp, low-pressure $f^n$ configuration.
Indeed, this phenomenon has already been experimentally observed in
resonant inelastic x-ray scattering measurements for both Ce and Pr.
\cite{rueff06,bradley12} As each
configuration carries its own moment, such variations would necessarily
modulate the $L\gamma_1'$ feature.

This potentially ambiguous result can be clarified by 
comparison to La, for which there is a true zero 4f 
occupancy.
The f-electron signature, Fig.~2(a), while reduced, does not fully
vanish in the collapsed-phase Ce spectrum, inconsistent with a complete
Mott delocalization. Pr NXES, which has a broader main $L\gamma_1$
peak than Ce, does not lend itself to a direct La comparison. The
persistence of 
its 4f hallmark will be demonstrated below.

Nd, which is not subject to any large VC transition, shows a minor
change in the $L\gamma_1'$ peak (Fig.~1). There is a
shift in $L\gamma_1'$ to a slightly higher energy ($\sim 0.3$ eV)
concomitant with a transition from an fcc to a distorted-fcc structure
at $\sim 18.0$ GPa.\cite{chesnut00,akella99} 
As the 4f electrons are still localized at
this pressure, the observed shift is likely due to subtle changes in the
relative positioning and subsequent electron transfer between conduction
subbands which are known to occur during the high-symmetry
transformations.\cite{velisavljevic05}
As will be shown shortly, however, the normalized
amplitude of the $L\gamma_1'$ peak is unchanged during this shift. We
note that this change is not associated with any known delocalization
transition; for example, prior diffraction and electrical resistivity
measurements suggest that 4f delocalization in Nd occurs gradually
beginning only at 100 GPa.\cite{velisavljevic05}


\underbar{Theoretical Predictions and Results:}
To this point we have made qualitative arguments regarding the
pressure dependence of the $L\gamma_1'$ shoulder. We
now supplement this with a theoretical treatment.
In the Kramers-Heisenberg formalism,
the NXES intensity is\cite{LuukRMP, KotaniRMP},

\begin{equation}
I_{g}(\omega) \propto \sum_{j} \biggr\lvert 
\sum_{i} \frac{\langle j | \hat{T} | i \rangle \langle i | \hat{a_{c}} 
| g \rangle}{ E_{j} - E_{i} - \omega -i \Gamma_{i}} \biggr\rvert ^{2},
\end{equation}
where $\hat{T}$  is the dipole operator for $4d\rightarrow 2p$ transitions, 
$\hat{a_c}$ is the annihilation
operator of the core electron, Â $\omega$ is the energy of emission, and 
$|g\rangle, |i \rangle$
and $|j\rangle$
are the ground,
intermediate and final states respectively with energies
$E_g, E_i$ and $E_j$.
To account for finite temperature,
Eq.~1 is modified assuming a Boltzmann
distribution:
\begin{align}
\langle I(\omega) \rangle _T = 
\sum_g e^{-E_g/kT} I_g(w) \,\big/\,
\sum_g e^{-E_g/kT} 
\end{align}
where $T$  is the working temperature (300 K in the analysis below) and $k$ is
the Boltzmann constant.

The electron states are determined by diagonalizing a Hamiltonian 
combining a single
impurity Anderson model \cite{anderson61,KotaniRMP} with 
interactions accounting for multiplet terms.
The pressure-dependent hybridization V is calculated by
a first-principle approach.\cite{SM} In Fig.~2, we
present the results of these calculations compared to the experimental
data. It must be noted that in the theoretical results, the atomic
multiplet features composing the $L\gamma_1'$ satellite are sharper than
they appear in experiment. This is a consequence  of using only five
discrete conduction orbitals in place of the true, broad 5d band for
hybridization. Such an approximation is necessary to ensure reasonable
computation time.  Nevertheless, it is clearly demonstrated that the
steplike decrease in $L\gamma_1'$ intensity observed in Ce and Pr
concomitant with VC is consistent with a sudden 
increase of 4f-conduction band hybridization as
predicted by the KVC model.

\begin{figure}
\centering
\includegraphics[width=2.4in]{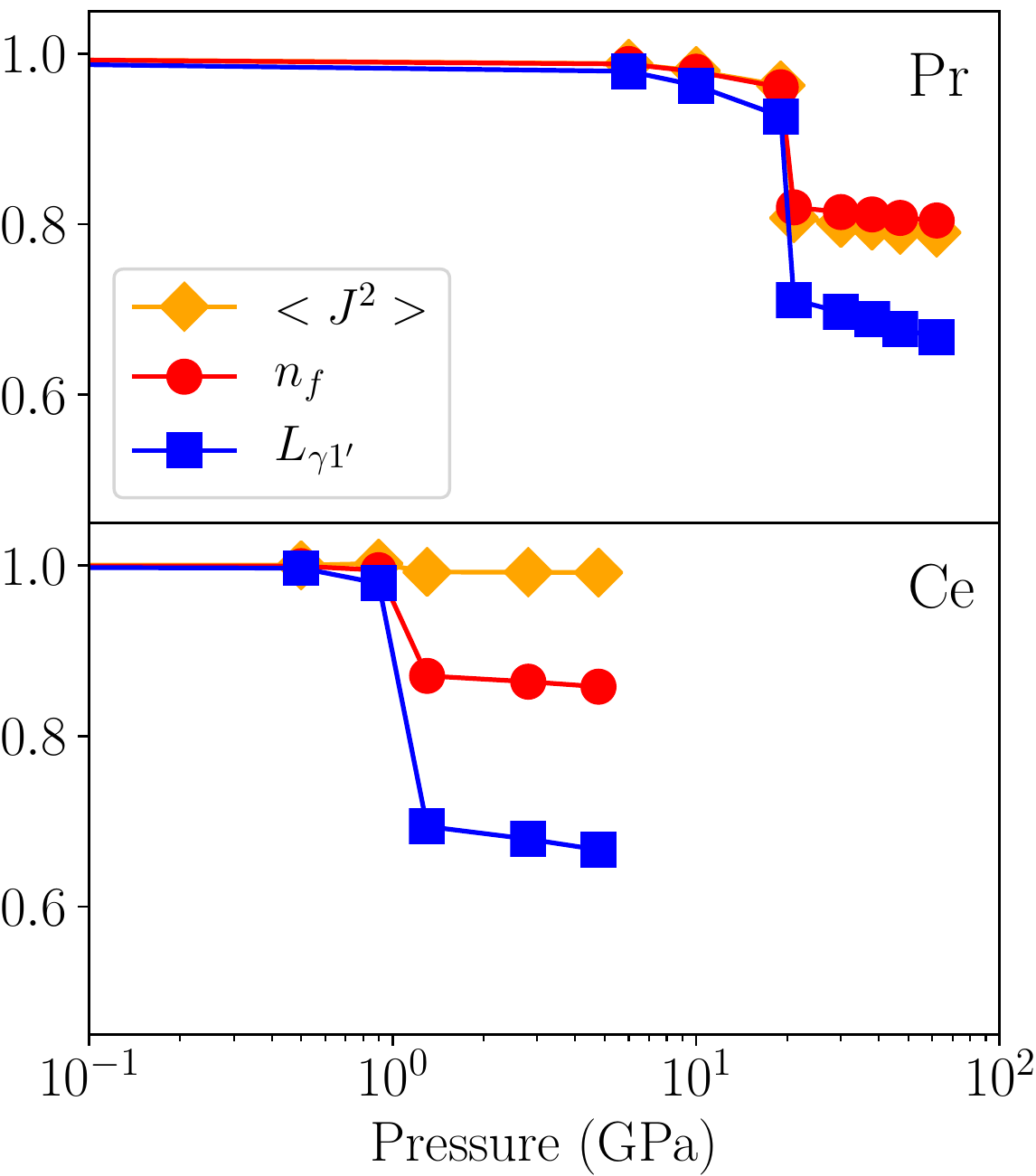}
\caption{(Color online.) The calculated bare magnetic
moments~($\langle J^2 \rangle$), 4f occupation numbers ($n_f$) and
shoulder peak intensities ($L_{\gamma1'}$) at various pressures for
cerium and praseodymium. All 
quantities are normalized to their zero-pressure value.}
\end{figure}

The physical observables are calculated by:
\begin{equation} 
A_T
=\frac{\sum_{g} e^{-E_g/kT} \langle g|\hat{A}|g \rangle}
{\sum_{g} e^{-E_g/kT}},
\end{equation} 
where
$\hat{A}=\hat{n_f}=\sum_{\nu} a^{\dagger}_{f,\nu}a_{f,\nu}$ for obtaining the 4f
occupation number and $\hat{A}=\hat{J^2}$ for the local 4f moment. The
values are normalized to zero pressure, as shown in Fig.~3.
The results demonstrate that both 4f occupancy and $L\gamma_1'$
intensity decrease as pressure goes up, with the $L\gamma_1'$ intensity
doing so at a slightly faster rate. The important point, however, is
that a persistent $L\gamma_1'$ feature is indicative of continued 4f
localization.

The local moment $\langle J^2\rangle$ behavior is more
complicated, since it depends on not only the 4f occupancy but also the
occupancy of each j level. Taking Ce first, there are j=5/2 and
j=7/2 states with spin-orbit coupling (SOC). The occupancy of the levels
will be a function of the 4f on-site energy, $\epsilon_f$ and the
hybridization V, as shown in Appendix A of \cite{GunnarssonPRB}. In the
V=0 limit, the system follows Hund's rule and only j=5/2 ($\langle
J^2\rangle$=8.75) is occupied. As V turns on, the occupancy of j=7/2 will
grow, and the ratio ($n_{7/2}/n_f$) depends on $\epsilon_f$: When
$\epsilon_f$ is deep in the valence band the ratio is small, so $\langle
J^2 \rangle$
decreases as $n_f\sim n_{5/2}$ decreases with rising V; In the other
limit, when $n_{7/2}/n_f$ reaches 8/14 at large V, $\langle J^2 \rangle$ increases as
V goes up.\cite{AKM05} For Ce, $\langle J^2\rangle$, Fig.~3(bottom), the
effects from decreasing $n_f$ and increasing $n_{7/2}/n_f$ compensate
each other.  $\langle J^2 \rangle$
stays almost constant as pressure goes up.
Ref.~\cite{lipp12} reached a similar conclusion concerning
the more paramount importance of $n_f$ in tracking the satellite
peak, ascribing the (somewhat larger) change reported
there in $\langle J^2 \rangle$ mostly to
the change in occupation, so that the latter is more fundamental.
Our conclusions are even somewhat more strong in implicating
the occupation, since $\langle J^2 \rangle$ is almost completely stable.

Figure 3 (top) similarly emphasizes an abrupt change in $4f$ occupation
also occurs through the Pr VC transition, concommitant with the satellite peak
evolution. 
$\langle J^2 \rangle$ is also reduced.  This is in contrast
to results reported for Pr in 
\cite{AKM05}.  The reason is the challenge
of considering
the full set of rotationally invariant Coulomb interactions
within dynamic mean field theory (DMFT). 
Reference \cite{AKM05} 
included SOC but only the direct Coulomb
interaction.
As a result, $\langle J^2 \rangle$ for Pr and even Nd behaved
similarly to Ce: $\langle J^2 \rangle$ increases when volume decreases,
because only a higher j=7/2 level is mixed in the ground state as
hybridization turns on. We include all interactions, so all multiplet states of Pr mix with
the Hund's rule ground state when V is nonzero. In the pressure range
of our calculation $\langle J^2 \rangle$ drops but retains $\sim$80\%
of its ambient value
across the VC transition. 
Reproducing the experimental features, both Ce and Pr
$L\gamma_1'$
intensities undergo a large, sudden drop with the VC. These reductions,
however, are incomplete with $\sim$ 70\% and $\sim$ 65\% for Ce and Pr
respectively. 


\noindent
\underbar{Conclusion:}
We have presented a high-quality dataset of high pressure
$L\gamma_1$ NXES useful for the characterization of bare 4f electron
evolution in the early light lanthanide metals. These data are
supported by state-of-the-art modified atomic calculations which extend
previous work performed on Ce alone, to Pr and Nd. A central
conclusion concerns the evolution of the 4f occupation.
There are increasing indications that the most unified picture of the
NXES spectra  for the light lanthanides and their compounds might be
provided by $n_f$ \cite{Lipp2016,Bianconi1988} rather than measures of
the 4f magnetism- the number of  Bohr magnetons, total angular momentum. 
Our calculations provide crucial evidence of a
clear relationship between $L\gamma_1'$ intensity and 4f-conduction band
hybridization. Thus, despite the importance of Kondo screening of the
moments in the volume collapse of Ce and Pr, the
interpretation of their NXES spectra appears also to fit in a broader,
common picture focused on 4f occupation.


\section*{Acknowledgements}

NXES studies were performed under the auspices of the US Department of Energy by Lawrence Livermore National Laboratory under Contract No. DE-AC52-07NA27344 and at HPCAT (Sector 16), Advanced Photon Source (APS), Argonne National Laboratory. HPCAT operations are supported by DOE-NNSA under Award No. DE-NA0001974 and DOE-BES under Award No. DE-FG02-99ER45775, with partial instrumentation funding by NSF.  The Advanced Photon Source is a U.S. Department of Energy (DOE) Office of Science User Facility operated for the DOE Office of Science by Argonne National Laboratory under Contract No. DE-AC02-06CH11357. The theoretical work of W.-T. C. and R.T.S. was supported by the SSAA program under grant {DE-NA0002908}. Portions of the computational work were performed using the resources of the National Energy Research Scientific Computing Center supported by the U.S. Department of Energy, Office of Science, under Contract No. DE-AC02-05CH11231. C.J.J., B.M., and T.P.D. at SLAC are supported by the U.S. Department of Energy, Office of Basic Energy Sciences, Materials Sciences and Engineering Division, under Contract No. DE-ACO2-76SF00515 for the atomic multiplet calculations and interpretation. D.R.M. and G.T.S. acknowledge support from the United States Department of Energy, Office of Basic Energy Sciences, under grant {DE-SC0002194} and also by the Office of Science, Fusion Energy Sciences, under grant {DE-SC0016251}. G.R. and S.Y.S. were supported by the US National Science Foundation Grant DMR-1411336 (S.Y.S).

\end{document}


\beginsupplement
\title{Supplemental Materials for \\
"Pressure Effects on the 4f Electronic Structure of Light Lanthanides"}

\maketitle

\noindent
\section{Experimental Details}
All measurements were performed at beam line 16ID-D of the Advanced
Photon Source.  Samples (99.9\% purity) were loaded into panoramic
diamond anvil cells fitted with either 300- 700 $\mu m$ culet diamonds
for lower pressures or 100 $\mu m$ flat beveled diamonds for higher
pressures. To prevent oxide formation each sample was loaded in a glove
box flushed with dry nitrogen and brought up to a low pressure to ensure
a good seal.  Pressure was measured by optical fluorescence from ruby
chips loaded with the sample. The incident beam (11.3-20.0 keV) was
brought through the diamond with the measured fluorescence escaping
perpendicularly through the Be gasket, measured in the scattering plane.
La, Pr, and Nd NXES experiments were conducted with newly developed
polycapillary-based Bragg spectrometers.
\cite{heald12,mortensen13}
These spectrometers
use half-lens collimating optics coupled to flat crystal analyzers (Ge
333, Ge 440 and Si 440 for La, Pr, and Nd respectively) with either a
scintillation (Cyberstar) or 2D pixel detector (Dectris). All Ce
$L\gamma_1$ have been directly reproduced from Lipp {\it et al.}
\cite{lipp12}.  All spectra were normalized 
by incident flux and corrected for a linear 
background.

\noindent
\section{Theory Details}
The first piece of the Hamiltonian,
whose eigenvalues and eigenvectors determine the NXES intensity, Eq.~1,
is an Anderson impurity model,
\begin{equation} 
\begin{split} 
H_{1} &= \sum_{k,
\mu}\epsilon_{k}a^{\dagger}_{k, \mu} a^{\phantom{\dagger}}_{k,
\mu}+\epsilon_{f}\sum_{\nu}a^{\dagger}_{f,\nu}a^{\phantom{\dagger}}_{f,\nu} \\ &
+\epsilon_{p}\sum_{\zeta}a^{\dagger}_{p,\zeta}a^{\phantom{\dagger}}_{p,\zeta}+\epsilon_{d}\sum_{\eta}a^{\dagger}_{d,\eta}a^{\phantom{\dagger}}_{d,\eta}
\\ & +\sum_{k,\nu}V_{f,k}
(a^{\dagger}_{f,\nu}a^{\phantom{\dagger}}_{k,\nu}+a^{\dagger}_{k,\nu}a^{\phantom{\dagger}}_{f,\nu}) \\ & +
U_{ff}\sum_{\nu >
\nu'}a^{\dagger}_{f,\nu}a^{\phantom{\dagger}}_{f,\nu}a^{\dagger}_{f,\nu'}a^{\phantom{\dagger}}_{f,\nu'} \\ &
-U_{fp}\sum_{\nu,\zeta}a^{\dagger}_{f,\nu}a^{\phantom{\dagger}}_{f,\nu}[1-a^{\dagger}_{p,\zeta}a^{\phantom{\dagger}}_{p,\zeta}]
\\ & -
U_{fd}\sum_{\nu,\eta}a^{\dagger}_{f,\nu}a^{\phantom{\dagger}}_{f,\nu}[1-a^{\dagger}_{d,\eta}a^{\phantom{\dagger}}_{d,\eta}].
\end{split} 
\end{equation}
Here $\epsilon_i$ is the on-site energy for orbital $i$,
$a_{i\sigma}^{\phantom{\dagger}} (a_{i\sigma}^\dagger)$ is the
annihilation (creation) operator for an electron of orbital angular
momentum $i$, with $\sigma$ serving as a combined index of the $m_l$ and
$m_s$ quantum numbers, $U_{ff}$ is the f-electron Coulomb potential, chosen to take the value $U_{ff}$ = 6 eV.
$U_{fp(d)}$ is the 2p (4d) core-hole potential and is set to $U_{fd(p)}$ = 10.5 eV(17.8 eV). 
We define the energy required
to promote a 4f electron to the bottom of conduction orbital to be $E_{\Delta}(n) = E(\ket{4f^{n-1}5d^1})-E(\ket{4f^n5d^0})$. $E_{\Delta}(1)$
= 1.3 (0.8) eV\cite{rueff06} for cerium and $E_{\Delta}(2)$ = 5.3 (4.6) eV for praseodymium below (above) the VC transition. For praseodymium the promotion energy includes double-counting energy.\cite{sawatzky94}
$V$ is
the hybridization between the f-electron and conduction orbitals under
pressure. For simplicity we took $V_{f,k}$ = $V$. The way $V$ is determined is
described below. We assume a 2 eV band width for the five discrete
conduction orbitals, with orbitals which span the band
evenly.\cite{tanaka95}  In the absence of hybridization, the ground
state consists entirely of the nominal $f^n$ occupation ($n =1,2$ for Ce
and Pr respectively) as discussed above. As pressure increases across
the transition, the ground state becomes a mix of $f^n$ and $f ^{n\pm
1}$ configurations. We do not consider the $4f^{n+1}$ configuration
since the strong 4f-4f coulomb interaction makes the energy of this
configuration much higher than the other two configurations- it will not
be prominent in the ground state.\cite{rueff06}

The second term, $H_2$, completes the electron-electron interactions by
accounting for multiplet interactions terms:
\begin{equation}
\begin{split}
H_{2} & = \sum_{\nu_{1},\nu_{2},\nu_{3},\nu_{4}} g_{ff}  (
\nu_{1},\nu_{2},\nu_{3},\nu_{4}
)a^{\dagger}_{f,\nu_{1}}a^{\dagger}_{f,\nu_{2}}a^{\phantom{\dagger}}_{f,\nu_{3}}a^{\phantom{\dagger}}_{f,\nu_{4}}
\\ & 
+\sum_{\nu_{1},\nu_{2},\zeta_{1},\zeta_{2}} g_{fp}  (
\nu_{1},\zeta_{1},\zeta_{2},\nu_{2}
)a^{\dagger}_{f,\nu_{1}}a^{\dagger}_{p,\zeta_{1}}a^{\phantom{\dagger}}_{p,\zeta_{2}}a^{\phantom{\dagger}}_{f,\nu_{2}}
\\ & 
+\sum_{\nu_{1},\nu_{2},\eta_{1},\eta_{2}} g_{fd}  (
\nu_{1},\eta_{1},\eta_{2},\nu_{2}
)a^{\dagger}_{f,\nu_{1}}a^{\dagger}_{d,\eta_{1}}a^{\phantom{\dagger}}_{d,\eta_{2}}a^{\phantom{\dagger}}_{f,\nu_{2}}
\\ & 
+
\sum_{\nu_{1},\nu_{2}}\alpha_{\nu_{1},\nu_{2}}a^{\dagger}_{f,\nu_{1}}a^{\phantom{\dagger}}_{f,\nu_{2}}
+
\sum_{\zeta_{1},\zeta_{2}}\beta_{\zeta_{1},\zeta_{2}}a^{\dagger}_{p,\zeta_{1}}a^{\phantom{\dagger}}_{p,\zeta_{2}}
\\ &
+\sum_{\eta_{1},\eta_{2}}\gamma_{\eta_{1},\eta_{2}}a^{\dagger}_{d,\eta_{1}}a^{\phantom{\dagger}}_{d,\eta_{2}}.
\end{split}
\end{equation}
where  $g$ are the parameters of the Slater-Koster integrals, and 
$\alpha, \beta$ and $\gamma$
are the spin-orbit
coupling parameters for respectively the $f, p$, and $d$ orbitals. All of
these values (including the on-site energy $\epsilon_i$) were obtained
from Cowan's atomic structure code\cite{cowan81} 
using the Hartree-Fock approximation and including relativistic effects.
The values of $g$ are scaled by 0.8 to account for screening, except for the
exchange integrals of 4f and 4d orbitals, $G(4d, 4f)$, which we take 0.72
of the nominal value. The scaling factors are empirical for Ce, Pr, and
Nd, and give the correct shoulder peak locations.

\begin{figure}
\centering
\includegraphics[width=3in]{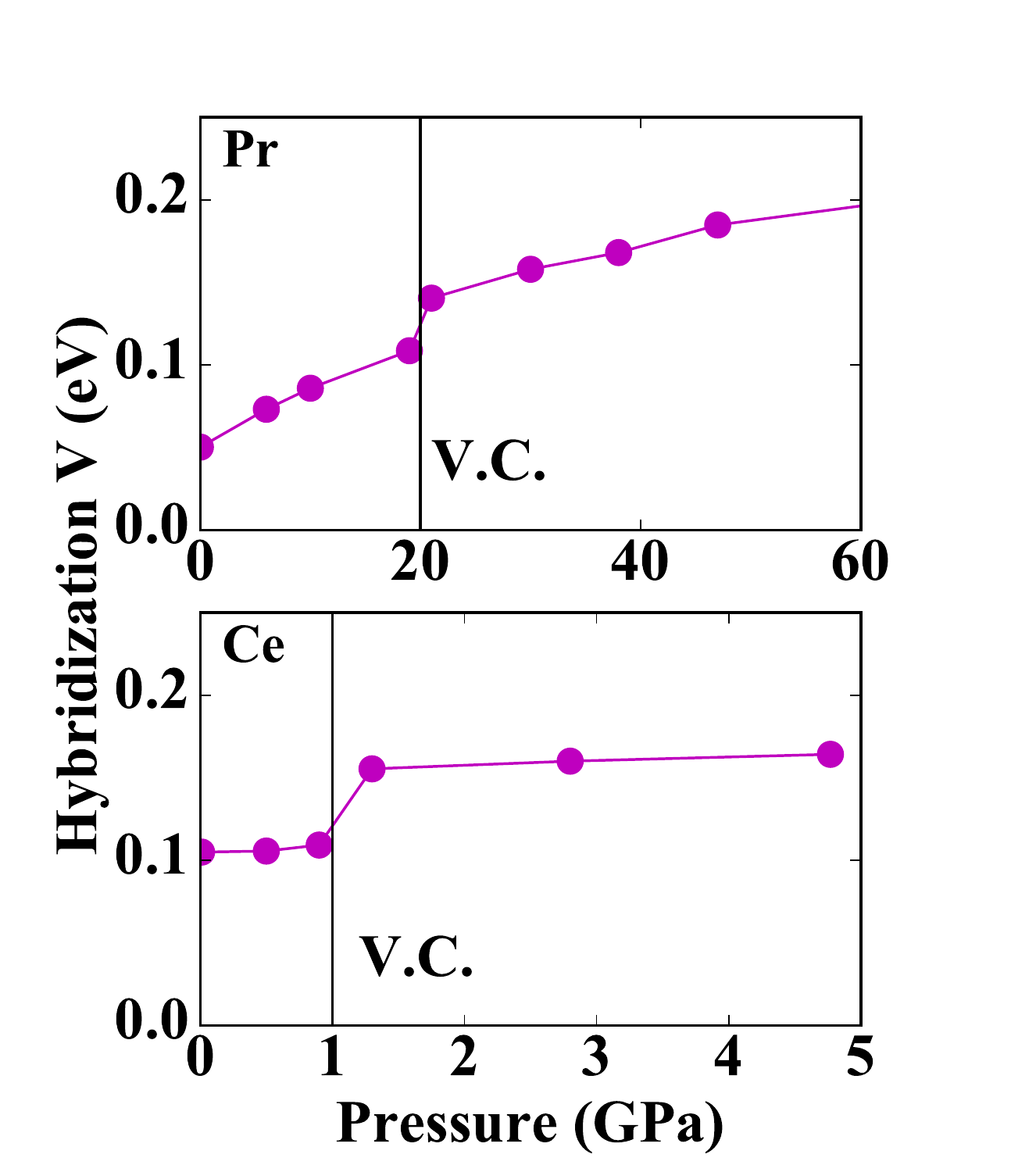}
\caption{(Color online)The hybridization V between
localized 4f orbitals and conduction bands for cerium and praseodymium
as a function of pressure. The vertical black lines indicate roughly the
pressures crossing the volume collapse transition at room temperature.}
\end{figure}

To obtain the hybridization matrix elements ($V$), we calculate the average
hybridization function $\Delta$ between the 4f and the
conduction states~\cite{Gunnarsson89, SavrasovPRB08, ShimPRL08,
ToropovaPRB07, HaulePRB10}.  
We first use the software routine developed for DFT+DMFT calculations by
Haule {\it et al.}~\cite{HaulePRB10} to compute the hybridization function
$\Delta_{\tau}$ in the absence of correlation effects (U=J=0)
for each atomic site ($\tau$) within the unit cell.
For the DFT part of the calculation, a mesh with 5000 k-points within the
Brillouin zone, plane-wave cutoff of $\mathrm{R_{MT}K_{max}}=10$ and maximum
angular momentum of $\mathrm{L_{MAX}}=12$ were used.  All calculations were
performed with the Perdew-Burke-Ernzerhof generalized gradient approximation
(PBE-GGA) for the exchange-correlation energy and $\mathrm{G_{MAX}}=14$.
Spin-orbit coupling was included using an extended energy range of
$\mathrm{E_{MIN}}=-10.0$ and $\mathrm{E_{MAX}}=10.0$.  
The lattice constants for the DFT calculation were determined using the
relation between volume and pressure for cerium and praseodymium taken
from~\cite{mcmahan98,Olsen85,Baer03}.

To compute the hybridization function, the routine first projects the DFT
eigenstates ($E(\vec{k},n)$, $\rvert \vec{k},n \rangle$) onto the orbital
basis for the impurity problem ($\rvert \tau,\nu \rangle$, $\nu = \{j,m_j\}$ ) and determines
the local lattice Green's function, which in diagonal basis is given by,
\begin{equation}
    G_\tau^{loc} (\nu;\omega) = \sum_{\vec{k}.n}
    \frac{
        \langle \tau,\nu \lvert \vec{k},n \rangle 
        \langle \vec{k},n \rvert \tau, \nu \rangle
    }{\omega  + \mu - E(\vec{k},n) - \Sigma(\vec{k},n;\omega)},
\end{equation}
where $\Sigma(\vec{k},n;\omega)$ is the DMFT embedded self-energy which in the
absence of correlations is zero.
The hybridization function $\Delta_\tau(\nu;\omega)$, is then calculated
using the DMFT self-consistency condition that the impurity Green's function
is equivalent to the local lattice Green's function,
\begin{equation}
    \frac{1}{\omega - E_\tau^{imp}(\nu) - \Sigma_\tau(\nu;\omega) - \Delta_\tau(\nu;\omega)} 
    = G_\tau^{loc}(\nu;\omega),
\end{equation}
where $E^{imp}_\tau(\nu)$ are the impurity energy levels and
$\Sigma_\tau(\nu;\omega)$ is the self-energy in the orbital basis for the
impurity problem which is likewise zero in the absence of correlations.
The hybridization function was calculated on the real-axis for each unique atomic 
site using relativistic harmonics for the orbital basis of 
the embedded impurity problem. The average hybridization function
$\Delta(\nu;E_F)$ was then determined by averaging the hybridization
function of all atoms within the unit cell,      
\begin{equation}
    \Delta(\nu;E_F) = \frac{1}{N}\sum_{\tau=1}^N \Delta_\tau(\nu;E_F).
\end{equation}
Finally the hybridization matrix elements ($V$) are estimated by
taking the square root of the imaginary part of the hybridization
function at the Fermi energy, $E_F$, with 
normalization:\cite{SavrasovPRB08, SWPR66}
\begin{equation}
|V|^2 = -\frac{1}{\pi}Im\sum_\nu \frac{\Delta(\nu;E_F)}{ 14\times N(E_F)}.
\end{equation}
Here $N(E_F)$ is the number of
states per atom in a unit cell. Once V is obtained, it can be inserted
into the single impurity Anderson model.